\documentclass[amsmath,amssymb,prl,hyperlink,twocolumn]{revtex4}

	\usepackage{graphicx}
	\usepackage{soul}
	\usepackage[colorlinks=true,citecolor=blue,linkcolor=magenta]{hyperref}
	\usepackage[usenames]{color}
	\usepackage{amsfonts}
	\usepackage{color}
	\usepackage{booktabs}
	\usepackage{multirow}
	\usepackage{float}
    \usepackage{times}
	\usepackage{upgreek}
    \usepackage[english]{babel}
     \usepackage{float}

\begin{document}

\title{Ruling out real-valued standard formalism of quantum theory}
\author{Ming-Cheng Chen$^{1,2,*}$, Can Wang$^{1,2,*}$, Feng-Ming Liu$^{1,2,*}$, Jian-Wen Wang$^{1,2}$, Chong Ying$^{1,2}$, Zhong-Xia Shang$^{1,2}$, Yu-Lin Wu$^{1,2}$, Ming Gong$^{1,2}$, Hui Deng$^{1,2}$, Fu-Tian Liang$^{1,2}$, Qiang Zhang$^{1,2}$, Cheng-Zhi Peng$^{1,2}$, Xiao-Bo Zhu$^{1,2}$, Adan Cabello$^{3,4}$, Chao-Yang Lu$^{1,2,\dag}$, Jian-Wei Pan$^{1,2}$
\vspace{0.2cm}}

\affiliation{$^1$ Hefei National Laboratory for Physical Sciences at Microscale and Department of Modern Physics, University of Science and Technology of China, Hefei, Anhui 230026, China}
\affiliation{$^2$ CAS Centre for Excellence and Synergetic Innovation Centre in Quantum Information and Quantum Physics, University of Science and Technology of China, Shanghai 201315, China}
\affiliation{$^3$ Departamento de Física Aplicada II, Universidad de Sevilla, E-41012 Sevilla, Spain}
\affiliation{$^4$ Instituto Carlos Ide Fisica Teoricay Computacional, Universidad de Sevilla, Spain}
\affiliation{}
\date{\today}

\begin{abstract}
Standard quantum theory was formulated with complex-valued Schrödinger equations, wave functions, operators, and Hilbert spaces. Previous work attempted to simulate quantum systems using only real numbers by exploiting an enlarged Hilbert space. A fundamental question arises: are the complex numbers really necessary in the standard formalism of quantum theory? To answer this question, a quantum game has been developed to distinguish standard quantum theory from its real-number analogue by revealing a contradiction in the maximum game scores between a high-fidelity multi-qubit quantum experiment and players using only real-number quantum theory. Here, using superconducting qubits, we faithfully experimentally implement the quantum game based on entanglement swapping with a state-of-the-art fidelity of 0.952(1), which beats the real-number bound of 7.66 by 43 standard deviations. Our results disprove the real-number formulation and establish the indispensable role of complex numbers in the standard quantum theory.
\end{abstract}

\maketitle

\begin{figure*}[htbp]
\includegraphics[width=0.7\linewidth]{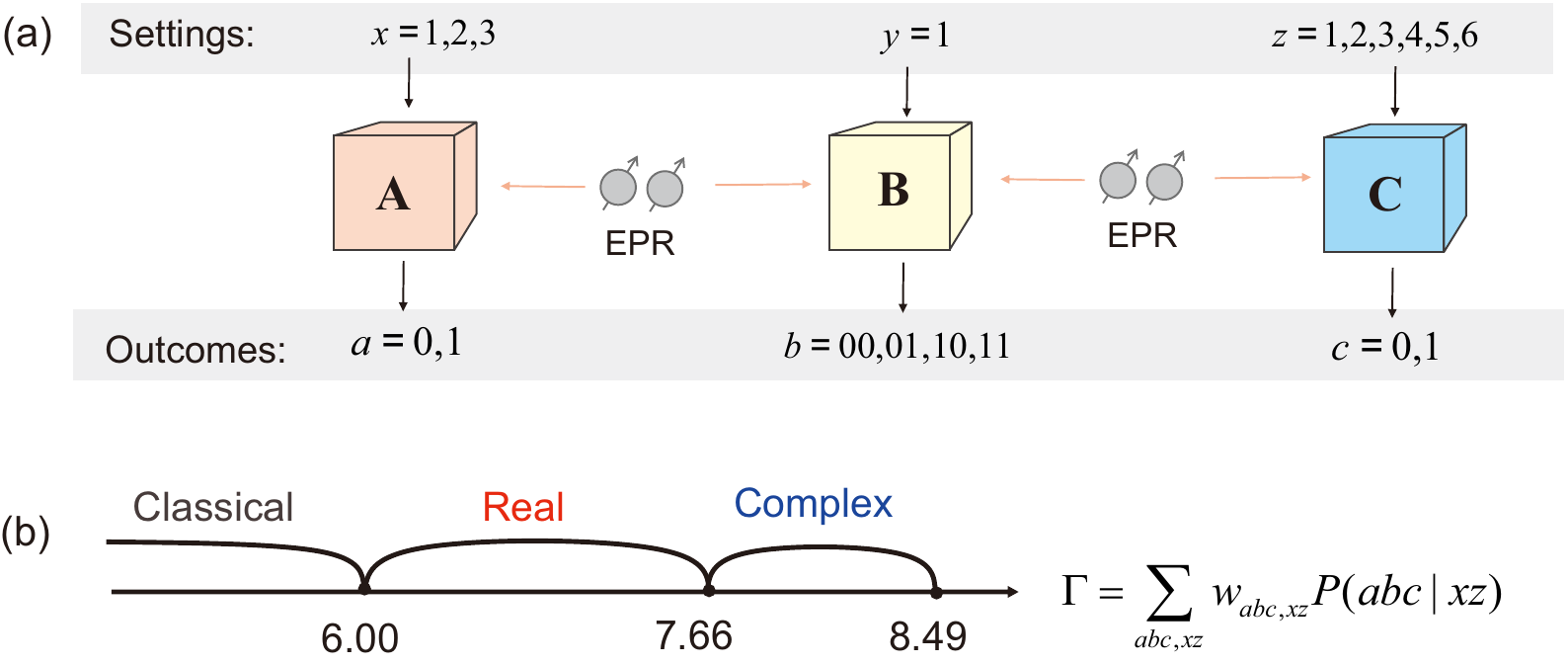}
\centering
\caption{A non-local game to disprove real-number model of quantum mechanics. (a) A three-party Bell-CHSH-like game based on entanglement swapping. Two EPR pairs are distributed between Alice and Bob, and between Bob and Charlie. The measurement settings are labelled as $x$, $y$, and $z$, and the outcomes are $a$, $b$, and $c$, respectively. Bob performs a complete Bell-state measurement. Alice and Charlies perform single-qubit local measurements. (b) The score of the non-local game divides the physics into three regimes: classical, real-number quantum mechanics, and complex quantum mechanics, which are experimental testable. The score $\Gamma $ is calculated from the weighted sum of the conditional joint probability distribution.}
\end{figure*}

Physicists use mathematics to describe nature. In classical physics, the real number appears complete to describe the physical reality in all classical phenomenon, whereas the complex number is only sometimes employed as a convenient mathematical tool. In quantum mechanics, the complex number was introduced as the first principle in Schrödinger's equation and Heisenberg’s commutation relation \cite{schrodinger1926undulatory0001, dirac1981principles0002}. The complex-valued wave function has been shown to represent the physical reality of quantum objects under certain physically plausible assumptions \cite{pusey2012reality0003}. Experimentally, the real and imaginary parts of the wave function have been directly measured \cite{lundeen2011direct0004}. Today, quantum mechanics with complex-valued wave functions seems the most successful theory to describe nature.

On the other hand, starting with von Neumann in 1936, many works \cite{birkhoff1936logic0005, stueckelberg1959field0006, stueckelberg1960quantum0007, guenin1961quantum0008, soler1995characterization0009, pal2008efficiency0010, mckague2009simulating0011, aleksandrova2013real0012, dyson1962threefold0013} have shown possible to universally simulate quantum systems using only real numbers by exploiting an enlarged Hilbert space in various alternative formalisms of quantum theory. For example, by adding an extra qubit $\left| \pm i \right\rangle =(\left| 0 \right\rangle \pm i\left| 1 \right\rangle )/\sqrt{2}$, a single-quantum system with a complex density matrix $\rho $ and Hermitian operator $H$ can be simulated through $tr(\rho H)=tr(\tilde{\rho }\tilde{H})$, where $\tilde{\rho }$ and $\tilde{H}$ are real and of the form:
\begin{align*}
\tilde{\rho }=(\rho \otimes \left| +i \right\rangle \left\langle  +i \right|+{{\rho }^{*}}\otimes \left| -i \right\rangle \left\langle  -i \right|)/2 \; ,\\
\tilde{H}=H\otimes \left| +i \right\rangle \left\langle  +i \right|+{{H}^{*}}\otimes \left| -i \right\rangle \left\langle  -i \right| \; .
\end{align*}
Therefore, it is interesting to ask a fundamental question why the complex number is necessary in the standard formalism of quantum theory. The standard quantum theory is established by the following four axioms \cite{nielsen2002quantum0014}:
\begin{itemize}
\item[(1)]
A pure quantum system is described by a unit complex vector in Hilbert space.
\item[(2)]
The state space of a composite quantum system is the tensor product of the state spaces of the component systems.
\item[(3)]
The dynamics of a closed quantum system is described by a unitary operator acting on the state vector.
\item[(4)]
A physical observable is described by a Hermitian operator and the measurement outcome obeys the Born rule.
\end{itemize}

In this work, we intend to investigate the real-number formalism of standard quantum theory, which satisfies the formalism in the above four axioms but replaces the complex vectors and operators in the Hilbert space by corresponding real vectors and operators, respectively. In this formalism, the dimension of real Hilbert space is not restricted to be the same as the complex Hilbert space. A distinguishing feature of the standard quantum theory from other quantum theories is at the second axiom, where the Hilbert space of composite quantum systems has a tensor-product structure.

Recently, Renou et al. developed an elegant scheme to provide an observable effect in quantum experiments to distinguish between the two formalisms \cite{renou2021quantum0015}. The scheme is a Bell-like \cite{bell1964einstein0016, clauser1969proposed0017, gisin2009bell0018, acin2016optimal0019, bowles2018self0020} three-party game based on deterministic entanglement swapping \cite{zukowski1993event0021}. The new theory predicts that the players obeying real-number formalism of standard quantum theory cannot obtain the maximal score that the standard complex-valued quantum theory allows, thus being falsified. An assumption used in the proposal \cite{renou2021quantum0015} is that the composite quantum state produced by two independent sources is the tensor product of the two independent quantum states, which has been ensured by the second axiom of standard quantum theory. Very recently, it is shown that the second axiom can be deduced from the first and fourth axioms, and is not independent \cite{carcassi2021four0022}. Specially, the tensor product structure of composite Hilbert space can be induced by experimentally accessible observables \cite{zanardi2004quantum0023}. 

Figure 1a shows an illustration of the three-party quantum game for Alice, Bob, and Charlie. First, two pairs of Einstein-Podolsky-Rosen (EPR) entangled qubits \cite{einstein1935can0024} are distributed between Alice and Bob, and between Bob and Charlie, respectively. Bob then performs a joint Bell-state measurement (BSM) on his two received qubits, which randomly projects them into one of the four Bell states:
\begin{align*}
  & \left| {{\phi }^{+}} \right\rangle \text{=(}\left| 0 \right\rangle \left| 0 \right\rangle +\left| 1 \right\rangle \left| 1 \right\rangle )/\sqrt{2}, \\ 
 & \left| {{\psi }^{+}} \right\rangle \text{=(}\left| 0 \right\rangle \left| 1 \right\rangle +\left| 1 \right\rangle \left| 0 \right\rangle )/\sqrt{2}, \\ 
 & \left| {{\phi }^{-}} \right\rangle \text{=(}\left| 0 \right\rangle \left| 0 \right\rangle -\left| 1 \right\rangle \left| 1 \right\rangle )/\sqrt{2}, \\ 
 & \left| {{\psi }^{-}} \right\rangle \text{=(}\left| 0 \right\rangle \left| 1 \right\rangle -\left| 1 \right\rangle \left| 0 \right\rangle )/\sqrt{2}. \\ 
\end{align*}
The four results are correspondingly registered as $b\in \{00,01,10,11\}$. Alice performs a single-qubit local measurement on her received qubit in the eigenstate bases of an operator ${{A}_{x}}$ ($x\in \{1,2,3\}$) which yields an outcome $a\in \{0,1\}$. Similarly, Charlie measures his qubit using operator ${{C}_{z}}$ ($z\in \{1,2,3,4,5,6\}$) and obtain his outcome $c\in \{0,1\}$. The score $\Gamma $ of the game is defined as the weighted sum of the conditional joint probability distribution $P(abc|xz)$, given by
\begin{align*}
\Gamma =\sum\limits_{abc,xz}{{{w}_{abc,xz}}P(abc|xz)},
\end{align*}
where ${{w}_{abc,xz}}\in \{-1,+1\}$ are the weights \cite{supplement}, $abc\in {{\{0,1\}}^{\otimes 4}}$ are the bit strings of measurement results, and $xz\in \{11,12,21,22,13,14,33,34,25,26,35,36\}$ are 12 combinations of local single-qubit measurement settings $x$ and $z$.

In the standard complex-valued quantum theory, we set Alice’s operator as ${{A}_{x}}\in \{Z,X,Y\}$, where $Z$, $X$, $Y$ are the standard Pauli matrices, and Charlie’s operator as ${{C}_{z}}\in \{Z\pm X,Z\pm Y,X\pm Y\}/\sqrt{2}$, where the score of the game reaches its maximum of $6\sqrt{2}$ ($\approx$8.49). However, if one assumes the players can only use real-number quantum resources (real-number states and real-number operations), the numerically found optimal score \cite{renou2021quantum0015} has an upper bound of 7.66, which sits between the classical limit of 6 and the quantum limit of $6\sqrt{2}$, as shown in Fig. 1b. This contradiction opens a way to a direct experimental test to distinguish between the complex-number and real-number representations of standard quantum theory.

\begin{figure*}[htbp]
\includegraphics[width=0.9\linewidth]{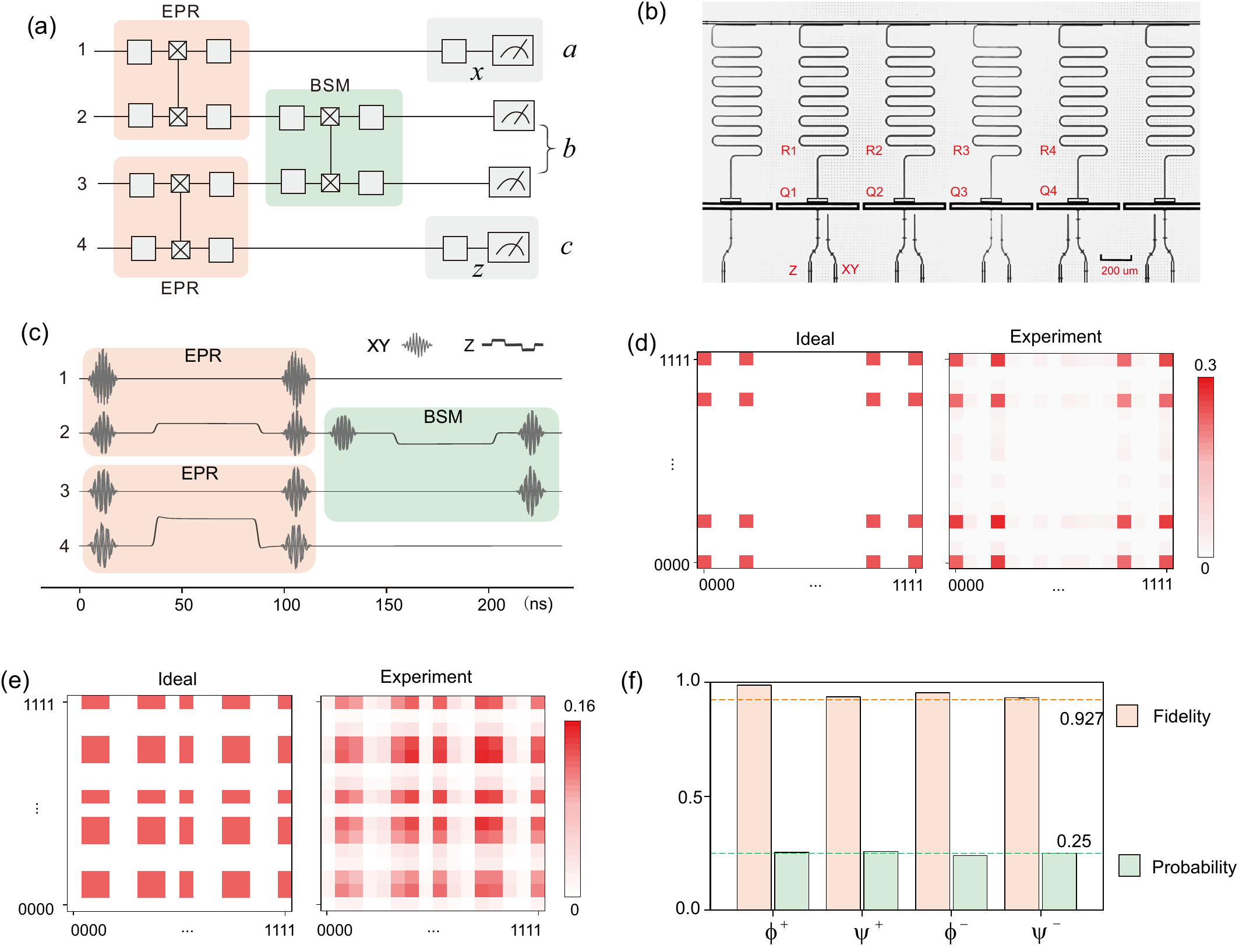}
\centering
\caption{Experimental implementation. (a) The quantum circuit used to perform the quantum game. The EPR pairs and BSM are realized by a two-qubit iSWAP operation with additional single-qubit gates. (b) Optical image of the superconducting quantum processor. The qubits (Q1-Q4) are placed in a linear array and have direct nearest-neighbour couplings. Each qubit has a XY driving and a Z tunning line and has an individual readout resonator (R1-R4) for measurement. (c) The experimental control pulses. Single qubit is rotated by Rabi resonance and two qubits are entangled by tunning the qubits into resonant spin exchange. Two EPR pulse sequences are used to prepare the two pairs of EPR entangled states. And the BSM pulse sequence is used to implement the joint Bell-state measurement. (d) The magnitude of density matrix of the product state of the two EPR pairs after the EPR pulses. The state fidelity is 97.5\%. (e) The magnitude of density matrix of the four-qubit entangled state after the BSM pulses. The state fidelity is 94.6\%. (f) The entangled states generated between Alice and Charlie conditional on the BSM outcomes. All are above 92.7\% threshold fidelity, and the average is 95.2\%.}
\end{figure*}

\begin{figure*}[htbp]
\includegraphics[width=0.7\linewidth]{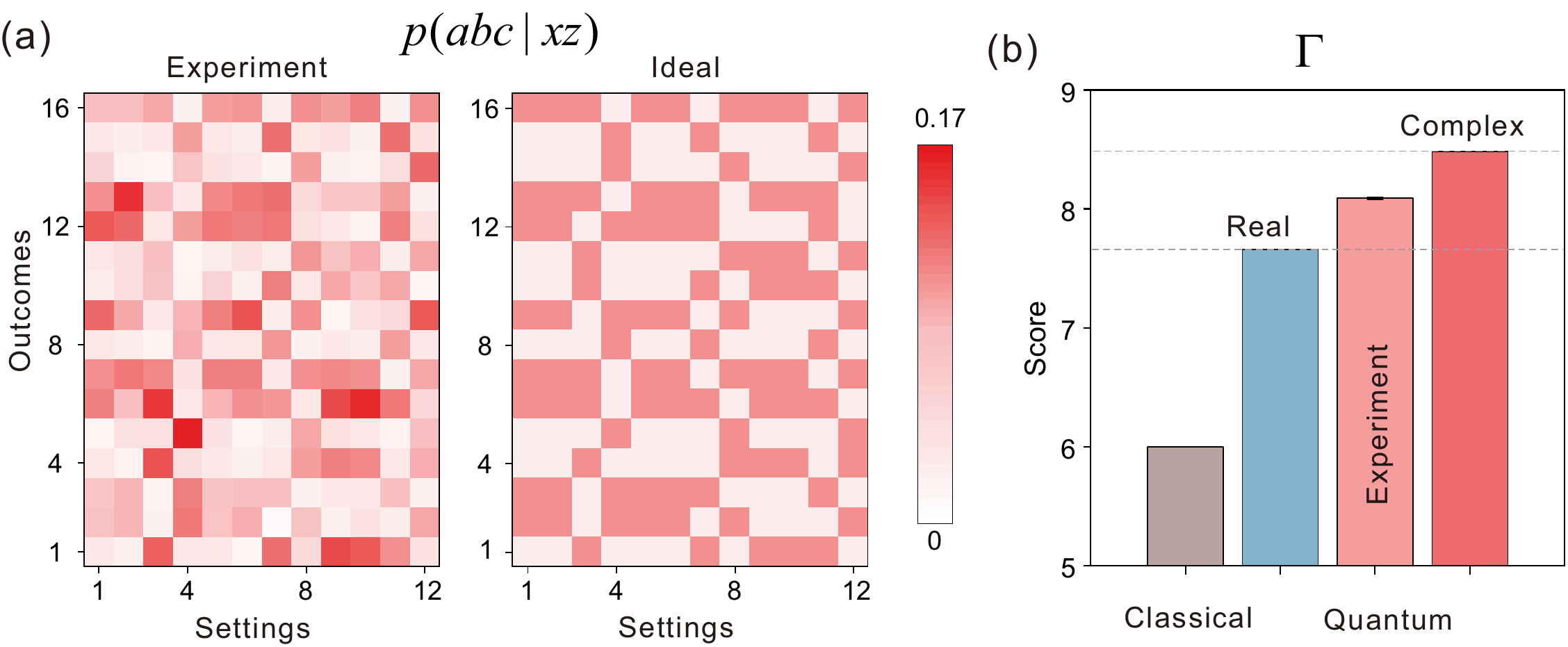}
\centering
\caption{Experimental results. (a) The conditional joint probability distribution matrices $P(abc|xz)$ from the quantum game (left) and the theoretical prediction from an ideal complex-number quantum theory (right). The row index represents the input measurement setting $\{xz\}$ and the column index is the output measured strings $\{abc\}$. (b) The obtained game score of the quantum experiment is 8.09(1), which is calculated from the experimentally measured conditional joint probability distribution matrices and is 43 standard deviations above the upper bound of the score (7.66) from ideal real-number quantum theory. The error bar on the experimental score presents one standard deviation. The score ($6\sqrt{2}$) from ideal complex-number quantum theory and score (6) from ideal classical hidden-variable theory are also shown.}
\end{figure*}

A generic quantum circuit for the experimental test is shown in Fig. 2a which involves three deterministic entangling gates for preparing two maximum entangled EPR pairs (1-2 and 3-4) and implementing the BSM between qubit 2 and 3. Very high quantum state and gate fidelities are required to surpass the real-number players \cite{renou2021quantum0015}. We design and fabricate a superconducting quantum processor (see Fig. 2b for its optical image) to implement the quantum game. To keep Alice’s and Charlie’s qubits (1 and 4) isolated from each other, we use I-shape transmon qubit design \cite{barends2013coherent0025} to increase the spacing of the qubits for reducing the non-neighbor coupling and with large frequency detuning, the maximum population exchange is smaller than ${{10}^{-9}}$ \cite{supplement}. The qubits are arranged in a linear array \cite{barends2014superconducting0026} and each has individual control and readout \cite{wallraff2004strong0027, jeffrey2014fast0028}. The qubits work at the transition frequency ${f}_{01}\!\sim\!5\text{ GHz}$ and are capacitively coupled to their nearest neighbors with $g/2\pi\!\sim\!5\text{ MHz}$.

Single qubit is controlled by microwave pulses on the Rabi driving (XY) line and fast flux-bias current on the frequency tuning (Z) line. Each qubit is dispersively coupled to a readout resonator (R1-R4) and is readout through a single coplanar waveguide. At the idle frequency, the qubits' average lifetime ${{T}_{1}}$ is measured to be 34.8 $\upmu$s and the average coherence time $T_{2}^{*}$ is 8.0 $\upmu$s. The qubits are initialized by idling 200 $\upmu$s to decay into the ground states. The average fidelity of single-qubit gates using 23 ns driving pulses is measured to be 99.88\%. The average error in the readout of the 16 computational bases after correcting the local bit-flip error is 0.34\%.

To implement the quantum circuit in Fig. 2(a), we design the control pulse sequences as shown in Fig. 2(c). The entangling interactions are in the form of iSWAP gates where the qubits fully exchange their populations as $\left| 01 \right\rangle \to i\left| 10 \right\rangle ,\left| 10 \right\rangle \to i\left| 01 \right\rangle$ in $\sim$50 ns by tuning two nearest-neighbor qubits into resonance \cite{barends2019diabatic0029, sung2021realization0030}. The quantum process fidelity of the iSWAP gate is measured to be 96.7\% with state preparation and measurement (SPAM) errors.

To characterize our experimental process, we make tomographic measurements of the quantum entangled states created after each step. First, after the first two EPR pulses (see Fig. 2c, at the point of $\sim$120 ns), we measure and reconstruct the density matrix of the product state of the two EPR pairs as shown in Fig. 2d. The experimental data agree well with the ideal case with fidelity of 0.975(1). Next, after the third entangling pulse (at the point of $\sim$230 ns in Fig. 2c), the four qubits become a fully entangled state. The measured density matrix in Fig. 2e shows a fidelity of 0.946(1) compared to the perfect state. Finally, projective measurements on the qubit 2 and 3 are performed, which swap the entanglement onto the independent and non-interacting qubit 1 and 4. Conditioned on random outcomes of the BSM, the qubits 1 and 4 are projected into one of the four Bell states correspondingly, whose quantum state fidelities are measured and plotted in Fig. 2e. The obtained average fidelity is 0.952(1), well above the threshold of \cite{renou2021quantum0015} $\sim$0.927.

Having verified the high quantum operation and state fidelities, we proceed to test the quantum game and measure the conditional joint probability distributions $P(abc|xz)$. Alice and Charlie’s qubits are first unitarily transformed using appropriate single-qubit gates \cite{supplement} and then measured in the computational bases. There are 12 different combinations of measurement settings and 16 possible measured bit strings. The main experimental results are plotted as a $12\times 16$ probability distribution matrix in Fig. 3(a) together with the ideal values. Summing over the obtained probability $P(abc|xz)$ with their corresponding weight \cite{supplement}, our quantum experiment gives a score of 8.09(1). As shown in Fig. 4b, the result violates the upper bound set by real-number quantum theory by 43 standard deviations, which provides strong evidence to disprove the real-valued standard formalism of quantum theory. Our implementation uses deterministic and high-fidelity quantum logic gates \cite{riebe2008deterministic0031, ning2019deterministic0032} and single-shot measurement with unity detection efficiency.
~\\

$^{*}$M. -C. Chen, C. Wang, and F. -M. Liu contributed equally to this work.
~\\

$^{\dag}$cylu@ustc.edu.cn

\bibliographystyle{apsrev4-1}
\bibliography{ref}

\end{document}